# Graphene on Silicon Modulators

V. Sorianello, G. Contestabile, and M. Romagnoli

*(Invited paper)*

*Abstract*— **Graphene is a 2D material with appealing electronic and optoelectronic properties. It is a zero-bandgap material with valence and conduction bands meeting in a single point (Dirac point) in the momentum space. Its conductivity can be changed by shifting the Fermi level energy via an external electric field. This important property determines broadband and tunable absorption at optical frequencies. Moreover, its conductivity is a complex quantity, i.e. Graphene exhibits both electro-absorption and electro-refraction tunability, and this is an intriguing property for photonic applications. For example, it can be combined as an active material for silicon waveguides to realize efficient detectors, switches and modulators. In this paper, we review our results in the field, focusing on graphene- based optical modulators integrated on Silicon photonic platforms. Results obtained in the fabrication of single- and double-layer capacitive modulators are reported showing intensity and phase modulation, resilience of the generated signals to chromatic dispersion because of proper signal chirp and operation up to 50 Gb/s.**

*Index Terms*— **Graphene Photonics, Integrated Optics, Modulators, Silicon Photonics.**

## I. INTRODUCTION

Graphene is an allotrope of Carbon with atoms arranged on a single layer, two-dimensional (2D), and hexagonal lattice [1]. The 2D symmetric atomic structure, made of covalent bonds, determines the peculiar electronic band dispersion with a linear energy-momentum relationship where the conduction and valence bands meet in single point (Dirac point) with no energy gap [2-4]. The peculiar band structure, together one-atom thickness, leads to intriguing electronic properties, as for example ultra-high mobility [5] and significant electric field effect. In fact, Graphene carrier concentration can be easily tuned, moving the chemical potential from the valence to the conduction band, by electrical gating [6].

The combination of zero-bandgap and electrically tuneable electronic properties is of particular interest in photonics, because it offers ultra-broadband tuneable optical absorption [7]. Indeed, its conductivity at optical frequencies is a complex quantity affecting both the optical absorption and the refractive index of the material [8-11]. This property makes Graphene an enabling material for realizing efficient devices for next generation photonics [12]. In particular, Graphene can be easily integrated on top of several active or passive photonic platforms such as silicon (Si), silicon nitride (SiN) or other dielectric waveguides in order to realize broadband photodetectors, fast electro-absorption and electro-refraction modulators.

In this paper, we review our recent results on Graphene-on-Si modulators, moving from the operation principle to the summary of the latest experimental demonstrations of modulators for telecom and datacom applications. Our work follows and improves the first demonstrations of Graphene on Si electro-absorption modulators (EAMs): in particular, the first EAM was based on a single layer of Graphene (SLG) transferred on top of a Si doped waveguide used to gate Graphene [13]. The device exhibited broadband spectrum of operation from 1.35µm to 1.6µm but with a limited bandwidth of 1 GHz. Few months later, the same authors demonstrated the first EAM based on two self-gating layers of Graphene (double layer Graphene, DLG) transferred on a passive Si waveguide [14], again with limited bandwidth (1 GHz). The first demonstration of an on-off keying (OOK) modulation at 10 Gb/s came few years later [15]. The device was based on a SLG transferred on top of Si doped waveguide. The same group improved recently their result showing up to 20 Gb/s operation with a similar device [16]. Larger bandwidth has been reported with DLG on SiN [17] and Si waveguides [18], with 30GHz and 35GHz electro-optical bandwidth, respectively.

The paper is organized as follows: SECTION II is dedicated to describe the Graphene optoelectronic properties, in SECTION III the Graphene on Si modulators concept and simulations are reported. In Section IV, our last results on SLG on Si modulators are reviewed. Section V shows the most promising results regarding DLG on Si modulators. Conclusions and a short final discussion are reported in Section VI.

## II. GRAPHENE OPTOELECTRONIC PROPERTIES

Being a 2D material, Graphene optical properties are properly modeled through its surface conductivity at optical frequencies $\sigma(\omega,\mu_c,\Gamma,T)$, where $\omega$ is the radian frequency, $\mu_c$ is the chemical potential, $\Gamma$ is a phenomenological scattering rate taking into account electron-disorder scattering processes, and $T$ is the temperature [10]. In the following, we consider a simplified scattering rate model with constant $\Gamma$ independent on the energy $E$ [10,19]. Under this approximation, the surface conductivity can be expressed through the Kubo Formula [19]:

V. Sorianello and M. Romagnoli are with Photonic Networks and Technologies Laboratory (PNTLab) at Consorzio Nazionale Interuniversitario per le Telecomunicazioni (CNIT), Via Moruzzi 1, 56124 Pisa, Italy. (e-mail: vito.sorianello@cnit.it and marco.romagnoli@cnit.it).

G. Contestabile is with TECIP Institute at Scuola Superiore Sant'Anna, Via Moruzzi 1, 56124 Pisa, Italy. (e-mail: giampiero.contestabile@santannapisa.it).
This work was supported by the European Union's Horizon 2020 research and innovation programme, GrapheneCore2 under grant 785219.

$$\sigma(\omega, \mu_c, \Gamma, T) =$$

$$\frac{ie^2(\omega+i2\Gamma)}{\pi\hbar^2}\left[\frac{1}{(\omega+i2\Gamma)^2}\int_0^\infty E\left(\frac{\partial f_d(E)}{\partial E} - \frac{\partial f_d(-E)}{\partial E}\right)\partial E - \int_0^\infty \frac{f_d(-E)-f_d(E)}{(\omega+i2\Gamma)^2-4(E/\hbar)^2}\partial E\right], \quad (1)$$

where $e$ is the electron charge, $\hbar$ is the reduced Planck constant, $E$ the energy, and $f_d(E)$ is the Fermi-Dirac distribution:

$$f_d(E) = \frac{1}{\left(e^{\frac{E-\mu_c}{k_BT}}+1\right)}, \quad (2)$$

where $k_B$ is the Boltzmann's constant. The first term in (1) is the contribution arising from intra-band electron-photon scattering processes, while the second is due to inter-band electron-photon scattering. The chemical potential $\mu_c$ is determined by the carrier density $n_s$ on the Graphene layer through [10]:

$$n_s = \frac{2}{\pi\hbar^2 v_F^2}\int_0^\infty E\left(f_d(E) - f_d(E+2\mu_c)\right)\partial E, \quad (3)$$

where $v_F\approx 9.5\cdot 10^5$ m/s is the Fermi velocity. Because of the pronounced field effect in Graphene, the carrier density can be easily controlled by application of a gate voltage and/or chemical doping, leading to a consequent significant tunability of the surface conductivity.

In order to have a clear visualization of the optical property of Graphene, the Drude model can be applied to the surface conductivity in order to derive an equivalent volumetric permittivity considering a finite thickness of the material $h_G$ [20]:

$$\varepsilon_\parallel = \varepsilon_r + \frac{i\sigma(\omega,\mu_c,\Gamma,T)}{\omega\varepsilon_0 h_G}$$

$$\varepsilon_\perp = \varepsilon_r, \quad (4)$$

where $\varepsilon_r$ is the permittivity of the surrounding medium and $\varepsilon_0$ is the vacuum permittivity. Figure 1 shows the calculated room temperature optical permittivity at 1550nm as a function of the Graphene chemical potential, for different scattering rate energies. Both the real and imaginary part vary significantly by changing the chemical potential, proving that both absorption and refraction are affected by the surface carrier concentration. The real part of the permittivity, mostly responsible for the refractive properties of the material, first gradually increases with the chemical potential, up to a maximum reached when the Pauli blocking condition is met ($\mu_c=\hbar\omega/2$), then decreases with almost linear slope. The imaginary part of the permittivity, mostly linked to the material absorption, is characterized by a first constant region where the chemical potential is below the Pauli blocking energy threshold ($\mu_c<\hbar\omega/2$) and the inter-band photon-electron scattering dominates, with charges of the valence band promoted to the conduction band. When the chemical potential is close to the Pauli blocking energy threshold ($\mu_c=\hbar\omega/2$), the imaginary part of the permittivity changes significantly with the chemical potential, here the inter-band and intra-band occurs simultaneously. Finally, at chemical potentials above the Pauli blocking energy threshold ($\mu_c>\hbar\omega/2$), the imaginary part of the permittivity is almost constant to a level that is determined by the intra-band photon-scattering energy. For this reason, the scattering rate is an important parameter for optoelectronic applications as it determines the maximum achievable transparency for the material, i.e. it influences fundamental figures of merit of modulators as insertion loss (IL) and extinction ratio (ER) [12]. There are many electron scattering mechanisms influencing the carrier relaxation time, i.e. scattering rate [3], and their discussion goes beyond the scope of the present paper.

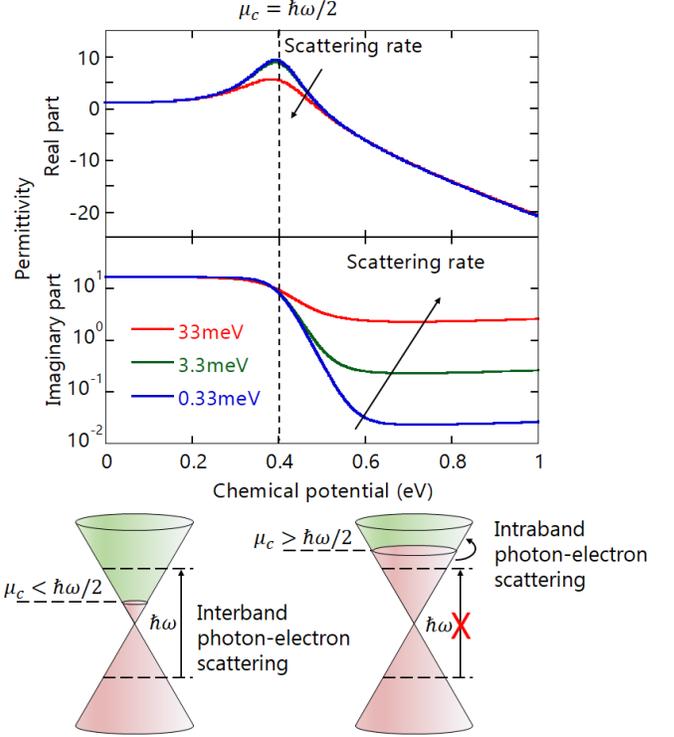

Fig. 1. Top panel: real and imaginary part of Graphene permittivity as a function of the chemical potential for different scattering rate energies. The permittivity is derived at room temperature, for a photon energy of 0.8eV (1550nm). Bottom panel: graphic illustration of the inter-band and intra-band photon-scattering processes.

However, it is important to highlight that the electron scattering affects the material mobility as well, which is a simple parameter to measure for the characterization of the material quality. In an ideal material the electron scattering would the lowest possible, which means in turn that the material mobility is the highest and scattering rate energy the lowest. The relation between the scattering relaxation time and mobility can be derived as in [12], to give a reference, the scattering energies used in fig. 1, 0.33 meV, 3.3 meV and 33 meV, correspond to mobility at $\mu_c$= 0.4eV of about 220 cm$^2$V$^{-1}$s$^{-1}$, 2200 cm$^2$V$^{-1}$s$^{-1}$ and 22000 cm$^2$V$^{-1}$s$^{-1}$. This means that Graphene modulators need high mobility to work efficiently.

### III. GRAPHENE ON SI MODULATORS: NUMERICAL SIMULATIONS

Graphene optical modulators exploiting the electric field effect can be realized with several geometries, here we are interested in dielectric waveguide modulators. Figure 2 shows the cross sections of the two possible Graphene-based integrated modulators: the SLG on Si doped waveguide (Fig. 2(a)), and the DLG on Si undoped waveguide (Fig. 2(b)).

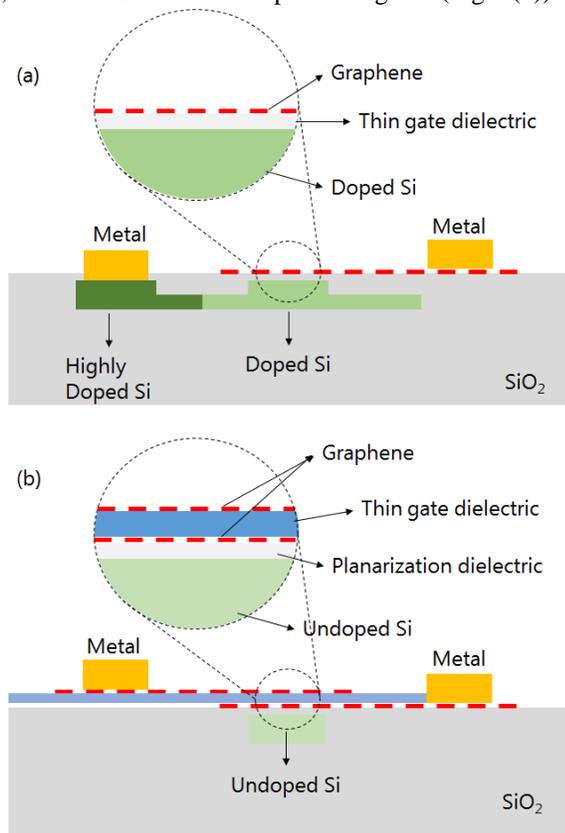

Fig. 2. Schematic cross section of: SLG on doped Si waveguide modulator (a), DLG on passive undoped Si waveguide modulator (b).

In both configurations, the waveguides are standard Silicon on Insulator (SOI) waveguides with a planarized oxide cladding thinned down to few nanometers from the top of the waveguide. Graphene optical properties are gated by means of the electric field generated in a capacitor, where Graphene is one or both the capacitor plates. By applying an external voltage to the capacitor, Graphene surface carrier concentration changes enabling modulation of the light guided in the waveguide.

In the SLG-based modulator, the capacitor consists of the Graphene-insulator-Si capacitor shown in the zoomed section of fig. 2(a). The thin planarized oxide on top of the waveguide act as the gate dielectric, while the Si ridge waveguide acts as the second plate of the capacitor. The last must be doped in order to reduce the series resistance of the device and to achieve high-speed modulation. In the DLG-based modulator, the capacitor is made of a stack two layers of Graphene separated by a thin dielectric film (zoomed section of fig. 2(b)). In this case, the waveguide is a fully passive channel waveguide so that modulators can be fabricated using any guiding material, e.g. undoped Si, Silicon Nitride (SiN), glass, etc.

The thickness of the dielectric insulator determines the modulator capacitance influencing both the modulator speed and the modulation efficiency, i.e. the amount of amplitude/phase change in the guided light versus applied voltage. A larger capacitance will have higher efficiency, because more charges will be moved on Graphene at same voltage, but lower speed, because of the higher RC constant.

We simulated the modulators reported in figure 2 around 1550nm with a commercial mode solver embedding Graphene optoelectronic model through the surface optical conductivity shown in (1) [21]. For the SLG device, the Si ridge waveguide is a transverse-electric (TE) single mode based on 220nm SOI platform with core width of 480 nm on top of a 60 nm thick Si slab. The top planarized oxide is a 5 nm thick silicon dioxide ($SiO_2$) layer. The Si waveguide is p-type doped with acceptor concentration of $5 \cdot 10^{17}$ cm$^{-3}$, with higher doping up to $10^{19}$ cm$^{-3}$ in the contact region. The distance between the highly doped region and the waveguide core was set to 500nm as a tradeoff between series resistance and optical losses. In the DLG device, two Graphene layers are placed on top of an undoped Si single mode TE waveguide with dimensions 480nm x 220nm. The waveguide has a planarized thin cladding of 5nm $SiO_2$. The two Graphene layers are separated by a 5 nm $SiO_2$ thick dielectric film.

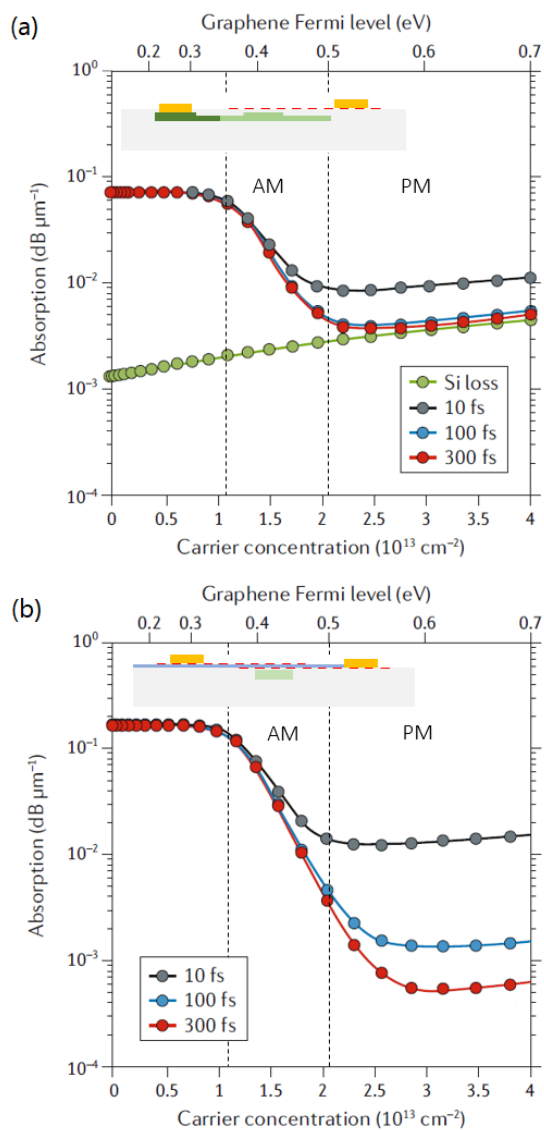

Fig. 3. Optical absorption as a function of the Graphene surface concentration for different scattering time constant: SLG on doped Si waveguide modulator (a), DLG on passive undoped Si waveguide modulator (b). [12]

Figure 3 shows the simulated optical absorption as a function of the carrier concentration and Fermi level (chemical potential

[3]) of Graphene for the SLG (fig. 3(a)) and DLG (fig. 3(b)) modulators, for three values of the scattering time: 10fs, 100fs and 300fs corresponding to mobility values of 220 cm$^2$V$^{-1}$s$^{-1}$, 2200 cm$^2$V$^{-1}$s$^{-1}$ and 6600 cm$^2$V$^{-1}$s$^{-1}$ at 0.4eV [12].

As anticipated in SECTION II, in both cases we observe two almost constant regions above and below the Pauli blocking condition (0.4eV at 1550nm), and a transition region across 0.4eV where the absorption changes significantly with the carrier concentration, i.e. applied voltage. This is the operating region where amplitude modulation (AM) is possible. The main difference between SLG and DLG modulators is in the achievable modulation depth of the two modulators. In fact, the DLG modulator exhibit a factor of two enhancement in the maximum absorption: ~0.12dB/µm with respect to ~0.06dB/µm of the SLG device. At the same time, also the transparency region above the Pauli blocking condition is different. In fact, we observe that the SLG modulator minimum absorption is strongly limited by the Si background loss due to free carrier effect in the doped waveguide (reported in green in fig. 3(a)) [22]. Conversely, the DLG modulator minimum absorption depends only on the Graphene quality, i.e. mobility. In case of a very high mobility a minimum loss below $10^{-3}$ dB/µm can be achieved. For this reason, Graphene DLG EAMs are very much promising in terms of attainable ER and IL.

In figure 1, we showed that while the imaginary part of Graphene permittivity saturate above the Pauli blocking condition, the real part still changes suggesting potential phase modulation. Figure 4 shows the change of the effective index of the SLG and DLG modulator, with mobility >2200 cm$^2$V$^{-1}$s$^{-1}$ (the real part of the permittivity does not change in the range, see fig. 1) as a function of the applied voltage and Graphene Fermi level energy.

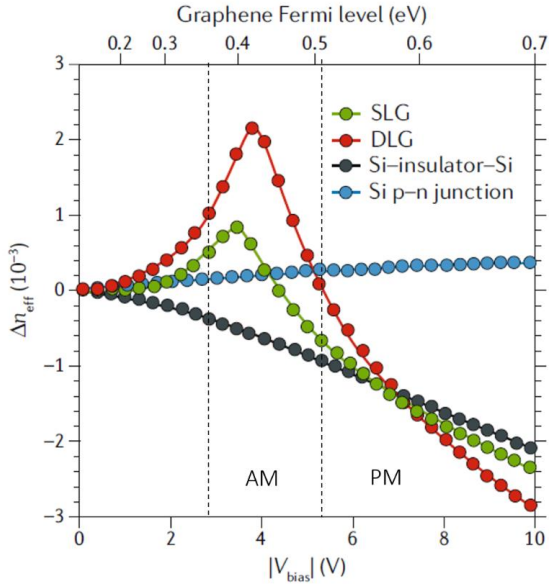

Fig. 4. Change of the effective index as a function of the Graphene Fermi level and applied voltage for: the SLG on doped Si waveguide modulator (green), the DLG on passive undoped Si waveguide modulator (red), Si capacitor modulator (grey) and Si pn junction modulator (light blue). [12]

In the same figure, the change in the effective index of standard Si pn junction and Si capacitor modulators are reported for comparison [12].

The applied voltage is calculated from the surface carrier concentration as follow [23]:

$$|V - V_{Dirac}| = \frac{e \cdot n_s}{C_{ox}} + K \frac{|\mu_c|}{e}, \qquad (5)$$

where $C_{ox}$ is the gate capacitance per unit area, $V_{DIRAC}$ is the voltage corresponding to the charge-neutral Dirac point, with K= 1 or 2 for SLG or DLG, respectively.

Figure 4 shows that both the SLG and DLG modulators potentially exhibit a maximum index change >4·10$^{-3}$, larger than Si modulators, both pn and capacitor based. The effective index always changes with the applied voltage without saturations, allowing phase modulation (PM) also in the region where the absorption does not change anymore. This operation region is suitable for the realization of phase modulators. In [12], we estimated phase modulation efficiencies of V$_\pi$L < 2.8 Vmm for SLG modulators and <1.6 Vmm for the DLG modulator, a factor of 5 improvement with respect to state of the art Si pn junction modulators [24] and Si capacitor modulators with same C$_{ox}$ [25]. An important figure of merit (FOM) for phase modulator is given by the product of the modulation efficiency V$_\pi$L multiplied by the propagation loss, i.e. FOM = V$_\pi$·IL. Here the material mobility plays a crucial role, as the FOM will be lower for high mobility Graphene. In [12] we estimate a FOM <2dBV when the mobility is larger than 2200 cm$^2$V$^{-1}$s$^{-1}$, i.e. more than 5 times better than any Si photonic modulator.

## IV. SLG ON SI MODULATORS

As anticipated in Section I, SLG EAMs have been reported in recent years working at 1, 10 and 20 Gb/s [13,15,16], being mainly bandwidth limited by the graphene-metal contact resistance. However, only a small attention has been given to consider the effects of the phase modulation in the device. In fact, even in EAMs, a certain amount of phase modulation is always present, as clearly noticeable in figure 4. This means that graphene-based EAMs generate chirped signals. In [26], we studied the chirp properties of the optical a signal at 1550 nm generated through a 100µm long SLG EAM with 5 GHz 3-dB electro-optic bandwidth, a modulation efficiency of 1dB/V, and a maximum attainable extinction ratio of 4.5 dB. The device was realized as described in the previous section, by transfer of Graphene on a planarized SOI ridge waveguide, details can be found in [26]. The device exhibited an IL of -19dB mainly due to the input/output grating couplers efficiency (5dB each), and to unexpected extra-losses of the air cladded Si waveguides before and after the device. The IL of the device was estimated by numerical simulation to be below 1dB [26]. We performed a direct measurement of the amplitude and phase of the output signal by using a complex spectrum analyzer. We drove the modulator with a sequence of isolated square pulses at 10 Gb/s with a 2.7V peak to peak driving voltage, biased above the Pauli blocking threshold ($\mu_c$>0.4eV). Figure 5 shows the measured amplitude and phase (figure 5(a)) of the optical signal and the corresponding instantaneous frequency (figure 5(b)). As expected from fig. 3 and fig. 4, the amplitude of the signal increases (the absorption decreases) while the phase decreases (index change decreases). Moreover, the phase follows linearly the amplitude causing a linear chirp of the signal. The bell shape of the pulse is due to the limited bandwidth of the device, which is not necessarily an impairment. Indeed, thanks to the electrical filtering effect, which gives a bell shape also to the phase profile, the associated instantaneous frequency chirp is linear and positive, i.e. linearly increases towards the trailing edge [27]. The linear positive chirp superimposed on the modulated signals is beneficial for chromatic dispersion compensation while transmitting the optical signal on standard single mode fibers (SMFs). Indeed, the Graphene modulator introduce a pre-chirp that compensate for fiber chromatic dispersion through the time lens effect [26]. Thanks to this effect, we demonstrated 100 km non-return to zero (NRZ) transmission on SMF at 10 Gb/s (figure 6) with a self-focusing distance (transmission distance for which there is the same signal sensitivity as in back-to-back) of 60 km [26]. The device was driven with a pseudo-random binary sequence (PRBS) ($2^{31}$-1-long) with $V_{pp}$=2.7 V driving voltage, corresponding to an effective ER of ~2.5dB.

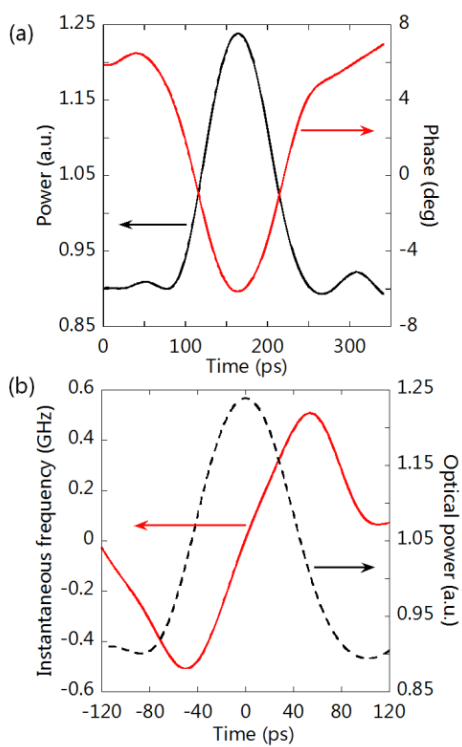

Fig. 5. Amplitude and phase versus time (a), and instantaneous frequency versus time of an optical pulse generated by a SLG on Si modulator (b) [26]

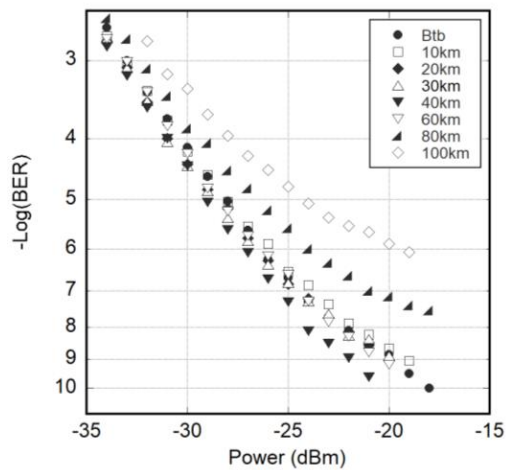

Fig. 6. BER vs. received power of a 10 Gb/s transmission up to 100 km SMF of a NRZ signal generated through a SLG on Si EAM.

The compactness of the device (100μm long) and the chromatic dispersion resilience can be combined to realize optical pre-emphasis obtained by cascading two EAMs [28]. Driving the cascaded circuit with data and properly attenuated and delayed inverted-data signals, the optical circuit behaves as a 1-tap filter frequency high-pass circuit. The cascaded EAM circuit introduce delay inverse weight compensation, i.e. the second modulator compensates for the slower part of the response of the first modulator (so enhancing the slow signal transitions due to the bandwidth limitation). This can be clearly appreciated in Fig. 7, where together with a picture of the cascaded modulators, a comparison of the eye diagrams in back to back is also reported. The evident eye opening gives a correspondent significant improvement in the sensitivity in back-to-back (6 dB) and in fiber transmission, in respect of the single graphene EAM [28].

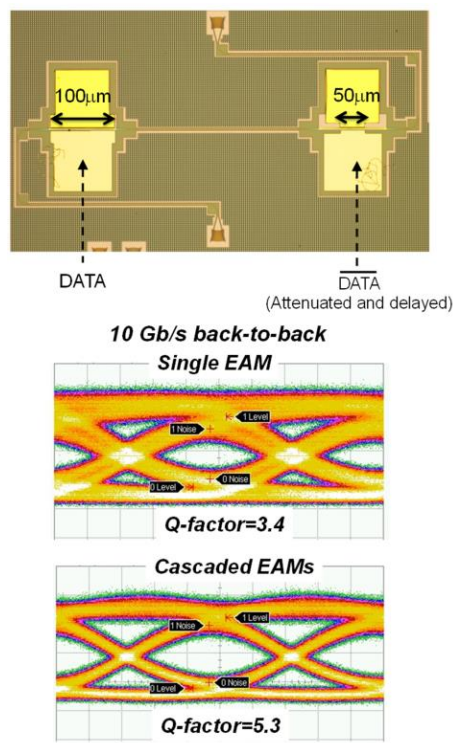

Fig. 7. Top panel: optical microscope image of the cascaded graphene EAMs. Bottom panel: comparison of 10Gb/s NRZ eye diagram of single and cascaded graphene EAM.

As mentioned in Section III, the phase response of Graphene modulator can be exploited to realize phase modulators. EAM and phase modulator are very similar devices, the main difference being the bias applied to the SLG capacitor, which, in the phase modulator case, must be well beyond the Pauli blocking threshold. The device operates in the 'transparency' region, where losses are only due to the waveguide propagation loss. In this regime, the phase modulation is dominant with an enhanced modulation efficiency compared to typical Si photonics modulators. Working on the electro-refraction effect of Graphene, we reported for the first time a Mach-Zehnder modulator based on two SLG on Si phase modulators [29]. The phase modulators were based on single mode TE ridge waveguides fabricated on a SOI platform. The Si waveguides were properly doped to allow for GHz bandwidth. The device is structurally similar to the EAM except for the length (300µm and 400µm); details are reported in [29]. The device exhibited a VπL = 2.8Vmm as expected form theory, even if we suffered higher insertion loss because of poor Graphene quality, i.e. mobility. The insertion loss of the device was about -27dB including the input/output grating couplers efficiency (5dB each), and unexpected extra-losses of the air cladded Si waveguides before and after the device. In addition, the phase modulator electro-optic bandwidth was limited by the metal–graphene contact resistance to 5 GHz. However, operation at 10 Gb/s was possible also obtaining SMF transmission up to 50 km (Figure 8) with dynamic ER of ~4dB.

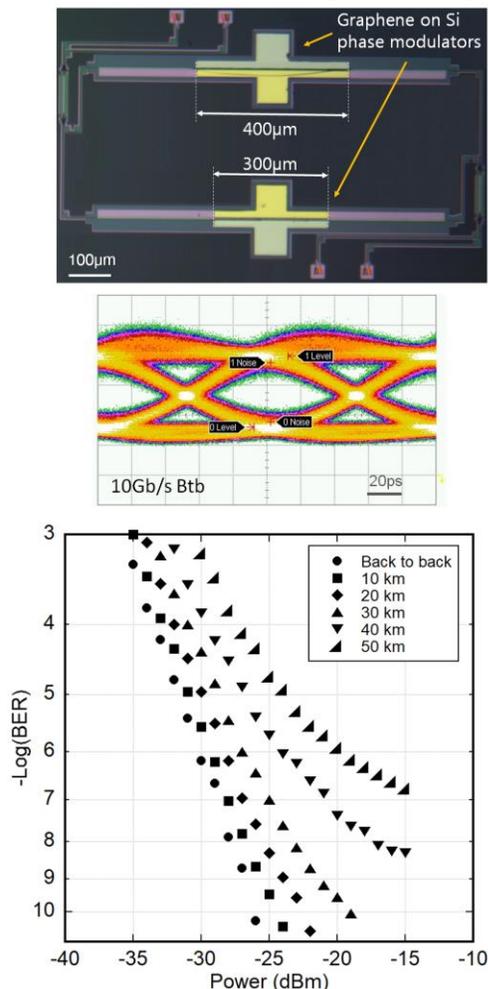

Fig. 8. Top panel: optical microscope image of the Graphene-based Mach-Zehnder modulator. Bottom panel: BER measurements of a $2^{31}-1$ PRBS NRZ signal at 10 Gb/s as a function of the received power in back-to-back configuration and after transmission over different standard SMF spool lengths.

In the previous examples of SLG on Si modulators the bandwidth was mainly limited by the contact resistance between the metal and Graphene. In particular, in all of these examples the contact was obtained by deposition of Palladium (Pd) on top of the Graphene layer, which is one of the material allowing good contact. However, we estimated a contact resistance as high as 1kΩ µm. The topic is object of many researches, which can be found in literature. Many different strategies have been developed to reduce the contact resistance including among others the use of different materials or particular geometries. Recently, it has been demonstrated that the use of Nickel (Ni) gives very low contact resistance (minimum theoretical 30 Ω µm) with respect to other materials [30]. A different and complementary strategy is the use of engineered patterning of the Graphene layer below the metal, which may improve the contact resistance (40 Ω µm at high electro-static doping) [31,32]. Finally, another strategy is the use of edge contacting in encapsulated Graphene stacks. In this case, Graphene is encapsulated with hexagonal Boron Nitride (hBN) forming a stack which is etched to allow metal contacting Graphene only at its edge. This approach has been demonstrated to allow very low contact resistance [33,34].

Adopting optimized contact strategies and engineered design, involving for example tight confinement in slot-waveguide, Graphene modulators can achieve modulation bandwidth above 100GHz [35-37]. The modulator bandwidth may be also enhanced by adopting the use of travelling-wave electrodes, as already demonstrated with Si and III-V modulators [38-39]. This approach has not been experimentally demonstrated with Graphene yet, the solution has been theoretically discussed and simulated recently in a Graphene on micro-fiber modulator claiming a bandwidth as high as 82GHz [40].

## V. DLG ON Si MODULATORS

DLG on Si modulators have two graphene layers acting as gates, as shown in fig. 1(b). This architecture allows larger electro-absorption and electro-refraction effect, which is approximately double with respect to the SLG case. Moreover, the self-gating of the two Graphene layers makes it possible to realize modulators on simple un-doped waveguides, not necessarily made of silicon. For these reasons, DLG on Si modulators are in principle more efficient and versatile with respect to SLG based modulators, as reported in Section III. However, the fabrication process of the stack of two layers of graphene separated by a high quality dielectric film is not trivial. In fact, the deposition techniques used for the dielectric thin film must satisfy strict specification in order to preserve the quality of the first Graphene layer, which is a pass-fail condition to realize efficient devices. For this reason, there are only few examples of Graphene modulators based on DLG on Si modulators operating at GHz bandwidth [14,17,18]. We recently demonstrated a DLG on Si EAM with 29 GHz bandwidth capable of NRZ modulation up to 50 Gb/s [30]. The modulator consists of a stack of two single crystals of Graphene grown by chemical vapour deposition (CVD) on Copper and transferred on the waveguide with a semi-dry method (details in [30]). The use of single crystals is of paramount importance to preserve the material high mobility. The dielectric gate is a 20nm thin film of SiN grown by plasma-enhanced CVD (PECVD) at 350°C [31]. This is about three times thinner than the alumina films used in [17] and six times thinner than [18]. This leads to a significant reduction of the operating voltages, i.e. larger modulation efficiency. Such improvement was possible because of a significant reduction of the contact resistance between metal and Graphene. We optimized the contact resistance down to ~300Ωµm at 0.2eV chemical potential. The combination of high quality material and reduced contact resistance allowed for the realization of the first DLG on Si EAM showing NRZ modulation eye diagram up to 50Gb/s (figure 9). The device exhibited a static ER of about 3dB and IL of ~20dB. The last was much larger than expected from simulation and from the overall quality of the graphene. We attributed the extra losses to the presence of residuals of metal and graphene on the input and output air-cladded waveguide sections. The measured eye diagrams of fig. 9 provided ER ranging from 1.5 dB at 25 Gb/s to 1.3 dB at 50 Gb/s. The reported eye diagrams are the state of the art for DLG on Si modulators.

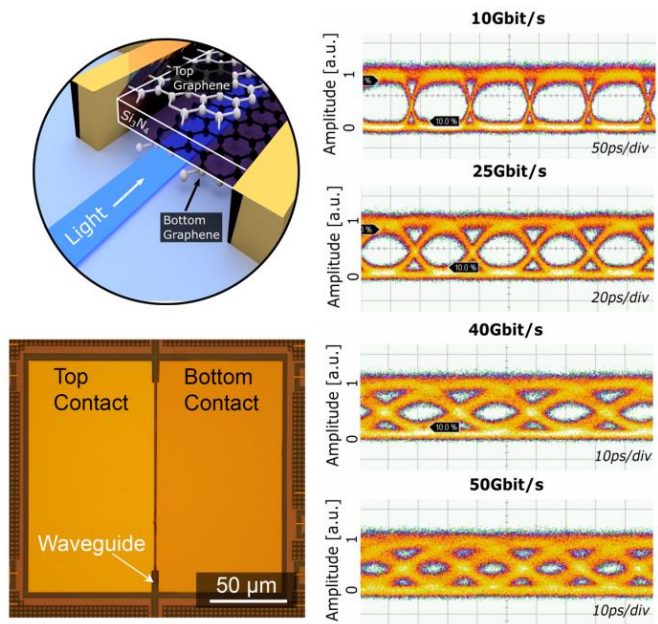

Fig. 9. Left panel: detail of the DLG on Si waveguide capacitor (top) and optical microscope picture of the fabricated device. Right panel: NRZ eye diagrams for a $2^{31}-1$ PRBS at data rates of 10Gb/s, 25Gb/s, 40Gb/s and 50Gb/s.

## VI. CONCLUSIONS

Graphene integrated photonics is attracting great interest for applications in telecom and datacom as a step forward in integrated photonics, in particular in order to improve Si photonics capabilities. Indeed, Graphene optoelectronic properties allows all the active functionalities: electro-absorption and electro-refraction modulation, photo-detection and, not mentioned in this paper, efficient thermal switching. In particular, DLG configurations permit the implementation of all these functionalities on almost any kind of passive waveguide, reducing, in principle, fabrication costs.

In this review paper, we discussed the optical modulation effects of Graphene integrated on silicon waveguides starting from an optoelectronic model of Graphene capable to show the operation principle of Graphene-based modulators. In particular, thanks to the enhanced electric field effect due to the 2D nature of the material, it is possible to exploit the efficient carrier accumulation on the Graphene layer for obtaining optical modulation in both the SLG and DLG configuration. In the second case, optical modulation is independent on the waveguide material, and shows a superior performance in terms of attainable ER and IL. All this considered together with the intrinsic high speed of the material, it is expected that Graphene modulators can provide the required bandwidth and efficiency to match the telecom and datacom roadmap evolution. In addition, the high efficiency of the modulation process allows for device miniaturization (<100µm for EAMs, <500µm for phase modulators) and reduced power consumption.

We also reviewed our work on both SLG and DLG on Si modulators carried out in the last years showing pure phase modulation, EAMs capable of generating signals with enhanced chromatic dispersion resilience and operation up to 50 Gb/s for the DLG EAMs.

Reported results demonstrate that Graphene photonics offers a combination of potential advantages in terms of both high-end performance and simplification in device manufacturing.

Concluding, we underline that the main material parameter for performance optimization is the carrier mobility as it affects significantly the overall modulator performance. In particular, a mobility >2200 $cm^2V^{-1}s^{-1}$ 0.4eV is required to realize competitive devices. These mobility values can be obtained by using single crystal graphene sheets and optimizing the transfer process, possibly exploiting a proper material encapsulation. To this extent, we expect a significant improvement of the technology in the next future, in order to satisfy the demands of performance of the next generation communication scenario. Further improvement of the modulation bandwidth towards 100GHz is required, as well as targeting other applications such as the coherent communication.

ACKNOWLEDGMENT

The Authors would like to acknowledge all coworkers at CNIT, Photonic Networks and Technologies Laboratory, TeCIP Institute of Scuola Superiore Sant'Anna, INPHOTEC Foundation and Graphene Labs, Italian Institute of Technology for the continuous support. All partners of the Graphene Flagship (WP8 Photonics and Optoelectronics) are also acknowledged for the continuous fruitful and collaborative work.

**Vito Sorianello** received the Ph.D. degree in electronic engineering from Roma Tre University, Rome, Italy, in 2010. In 2010, he received the IEEE Best Doctoral Thesis Award for his Ph.D. thesis on "Germanium-on-Silicon near-infrared photodetectors" by the IEEE Photonics Society Italian Chapter. He was a Postdoctoral Research Fellow with the Nonlinear Optics and OptoElectronics Lab of Roma Tre University, being involved and responsible for the fabrication and characterization processes of Germanium-on-Silicon photodetectors for three years. He is researcher at the Photonic Networks National Lab, Interuniversity National Consortium for Telecommunications, Pisa, since 2013. His main research interests include the modeling, design, and characterization of optoelectronic components and systems for the silicon photonics platform.

He is currently involved in several application-oriented research projects in collaboration with national and international research institutions and industries.

**Giampiero Contestabile** received the Laurea degree in Physics from "La Sapienza" University of Rome, Rome, Italy, and the Ph.D. degree in Electrical Engineering and Telecommunications from "Tor Vergata" University of Rome, in 1998 and 2001, respectively. From 1996 to 2000, he was with the Semiconductor Devices Group of "Fondazione Ugo

Bordoni," Rome. In 2001, he was with Optospeed Italia. Since 2002, he has been an Assistant professor with Scuola Superiore Sant'Anna, Pisa, Italy where, from 2018 he is an Associate Professor. He has co-authored more than 200 papers published in international peer-reviewed journals and presented in leading international conferences. His research interests include photonic integrated circuits, advanced optical systems, access networks, and semiconductor optical amplifiers and lasers

**Marco Romagnoli** received the Laurea degree in physics from the Universita` di Roma "La Sapienza," Rome , Italy, in 1982. He is currently the Head of Advanced Technologies for Photonic Integration at the Interuniversity National Consortium for Telecommunications and chief scientific officer, Pisa, Italy, a contract Professor at Scuola Superiore Sant'Anna, Pisa, and the former Director in R&D Department. In 1983, he started his activity at IBM Research Center, San Jose, CA, USA. In 1984, he joined Fondazione Ugo Bordoni in the Optical Communications Department working on optical components and transmission systems. In 1998, he joined Pirelli, where he was the Director of Design and Characterization and a Chief Scientist in R&D Photonics. In October 2010, he joined PhotonIC Corporation, a Si-Photonics company, as the Director of Boston Operations and the Program Manager at the Massachusetts Institute of Technology for the development of an optically interconnected multiprocessor Si chip. He is an author of more than 190 journal papers and conference contributions and a coinventor in more than 40 patents. He is in the technical committee of the major conferences in photonics such as CLEO/QELS, CLEO Europe, ECOC, MNE, Group IV Photonics, and served as an expert Evaluator for EC in the 6$^{th}$ Framework Program. He has more than 35 years of experience in the research field, especially in the area of photonic technologies for TLC.